\documentclass{aa}

\usepackage{graphicx}
\usepackage{txfonts}
\usepackage[table,xcdraw]{xcolor}
\usepackage{multirow}
\usepackage[table,xcdraw]{xcolor}

\usepackage{hyperref}
\hypersetup{colorlinks=true, citecolor=blue}

\usepackage{mhchem}
\usepackage{enumitem}

\newcommand{\kms}{km~s$^{-1}$}
\newcommand{\mvir}{M$_{200}$}
\newcommand{\rvir}{R$_{200}$}
\newcommand{\mstar}{M$_{*}$}
\newcommand{\Msun}{\rm M_{\sun}}
\newcommand{\reff}{R$_{\rm hm}$}
\newcommand{\ropt}{R$_{\rm opt}$}
\newcommand{\logten}{\ensuremath{\log_{10}}}
\newcommand{\cielo}{{\sc CIELO}}

\begin{document}

   \title{The \cielo~ project: The chemo-dynamical properties of galaxies and the cosmic web}
   
   \author{Patricia B. Tissera\inst{1,}\inst{2}\fnmsep\thanks{E-mail: patricia.tissera@uc.cl}
          \and
 Lucas Bignone\inst{3}
 \and
Jenny Gonzalez-Jara \inst{1,}\inst{2}
\and
Ignacio Mu\~noz-Escobar \inst{1,}\inst{2}
\and
Pedro Cataldi \inst{3}
\and
Valentina P. Miranda\inst{1,}\inst{2}
\and
Daniela Barrientos-Acevedo\inst{1,}\inst{2}
\and
Brian Tapia-Contreras\inst{1,}\inst{2} 
\and
Susana Pedrosa\inst{3} 
           \and
           Nelson Padilla\inst{4} 
           \and
           Rosa Dominguez-Tenreiro\inst{5,} \inst{6}
           \and
           Catalina Casanueva-Villarreal\inst{1,}\inst{2} \and
Emanuel Sillero\inst{1,}\inst{2} \and
Benjamin Silva-Mella\inst{1,}\inst{2} \and
Isha Shailesh\inst{1,}\inst{2} \and
Francisco Jara-Ferreira\inst{1,}\inst{2}
 }

        \institute{Instituto de Astrof\'isica, Pontificia Universidad Cat\'olica de Chile. Av. Vicu\~na Mackenna 4860, Santiago, Chile.
         \and
Centro de AstroIngenier\'ia, Pontificia Universidad Católica de Chile. Av. Vicu\~na Mackenna 4860, Santiago, Chile.
        \and
        Instituto de Astronom\'ia y F\'isica del Espacio, CONICET-UBA, Casilla de Correos 67, Suc. 28, 1428, Buenos Aires, Argentina.
        \and
        Instituto de Astronom\'ia Te\'orica y Experimental, Laprida 900, C\'ordoba, Argentina.
        \and
        Departamento de F\'isica Te\'orica, Universidad Autónoma de Madrid, E-28049 Cantoblanco, Madrid, Spain.
        \and
        Centro de Investigaci\'on Avanzada en F\'isica Fundamental, Universidad Aut\'onoma de Madrid, E-28049 Cantoblanco, Madrid, Spain
        }

   \date{Received October 10, 2024 ; accepted November 21, 2024}

  \abstract
    {
The \cielo~project introduces a novel set of chemo-dynamical zoom-in simulations, designed to simultaneously resolve galaxies and their nearby environments. The initial conditions (ICs) encompass a diverse range of cosmic structures, including local groups, filaments, voids, and walls, enabling a detailed exploration of galaxies within their broader cosmic web context.
    }
    {
This study aims to present the ICs and characterise the global properties of \cielo~galaxies and their environments. Specifically, it focuses on galaxies with stellar masses ranging from $10^8$ to $10^{11} \Msun$ and investigates key scaling relations, such as the mass-size relation, the Tully-Fisher relation (TFR), and the mass-metallicity relation (MZR) for both stars and star-forming gas. 
    }
    {
We employed the DisPerSe algorithm to determine the positions of \cielo~galaxies within the cosmic web, with a particular focus on the Pehuen haloes. The selection of Local Group (LG) type volumes was guided by criteria based on relative positions and velocities of the two primary galaxies. The Pehuen regions were selected to map walls, filaments and voids. Synthetic SDSS i, r, and g band images were generated using the SKIRT radiative transfer code. Furthermore, a dynamical decomposition was performed to classify galaxy morphologies into bulge, disc, and stellar halo components.
    }
    {
The \cielo~galaxies exhibit stellar-to-dark matter fractions consistent with both observational data and other simulation results. These galaxies align with expected scaling relations, such as the mass-size relation and TFR, indicating effective regulation of star formation and feedback processes. The mass-size relation reveals the expected dependence on galaxy morphology. The gas and stellar MZRs also agree well with observational data, with the stellar MZR displaying strong correlations with galaxy size (\reff) and star formation rate (SFR). This indicates that smaller, less star-forming galaxies tend to have higher metallicities. Future investigations will delve into the chemo-dynamical properties of bulges, discs, and stellar haloes, exploring their connections to assembly histories and positions within the cosmic web.
    }
    {}
    
   \keywords{ Galaxies: abundances - Galaxies: formation - Galaxies: Galaxy - Galaxies: evolution}

\maketitle

\section{Introduction}

In the current cosmological paradigm, galaxies reside within  dark matter haloes that evolved in a hierarchical fashion, from small structures to the largest ones, growing by smooth accretion  and mergers \citep{wr78, momaowhite1998}. As the large-scale structure takes form, a diversity of local environments arises,  shaping the evolution of haloes \citep{dressler1980}. 
Baryons, initially in the form of gas, cool and condense within dark matter haloes, where they are subsequently transformed into stars. The regulation of star formation in haloes of varying masses and at different redshifts remains a topic of debate, as multiple mechanisms across a wide dynamical range can influence the properties of the interstellar medium (ISM), thereby modifying the conditions required for star formation. Models and simulations suggest that  Supernova  and Active Galactic Nucleus (AGN) feedback play a critical role in modulating the transformation of gas into stars in galaxies of different masses \citep[e.g.][]{dekelsilk1986, whitefrenk1991,tiziandimatteo2005,oppenheimer2017, pillepich2019}. The interplay between internal physical processes and environmental factors is complex, and achieving a comprehensive understanding remains challenging.

As stars evolve, new chemical elements are synthetised and ejected into the ISM and even out of galaxies. These chemical elements enrich the material from which new stars are subsequently formed. As a result of the complex co-evolution between galaxies and their global environment, chemical abundances and patterns are shaped and modulated.\citep{maiolino2019, matteucci2021}. Thus, they could serve as fossil records from which significant events in the evolutionary history of galaxies could be inferred.\citep[e.g.][]{tins1979,matteucci1986}. The implementation of chemical evolution in cosmological galaxy formation codes is a powerful tool for studying chemical abundance patterns in galaxies and linking them to their formation history \citep[e.g.][]{mosconi2001,lia2002,koba2007}. 
Numerical studies have shown the impact of galaxy-galaxy interactions and mergers in triggering star formation and redistributing angular momentum \citep{perez2011,torrey2012,dimatteo2013, sillero2017,moreno2019}.  As enriched material is expelled by galactic outflows triggered by Supernova (SN) \citep[e.g.][]{oppenheimer2017} or AGN \citep[e.g.][and references therein]{wright2024} events or stripped out of galaxies by environment-dependent mechanisms,  they contribute to polluting the circumgalactic and intergalactic media \citep[e.g.][]{klimenko2023}. 
Previous studies have highlighted the impact of gas infall along filaments as a potential mechanism for fueling star formation in galaxies \citep{ceverino2016} and reshaping the distribution of chemical elements \citep{collacchioni2019}. The interplay with the environment could have affected the gas-richness and the metallicity distribution in galaxies, therebay impacting the metallicity gradients  \citep[e.g.][]{sanchezmenguiano2016,franchetto2021}. Numerical results showed the exchange of material with the Circumgalactic Medium (CGM), driven by SN feedback, which produces larger mass-loading factors in small galaxies \citep[e.g.][]{scan08,hopkins2013, vogelsberger2013,muratov2017}, which can vary with redsfhit as the physical conditions within galaxies are different \citep{bassini2024}. Recently, \citet{jara2024} showed a secondary dependence of the MZR on the merger history of galaxies by identifying that strong negative and positive metallicity gradients tend to be located at higher and lower enrichment level at a given stellar mass. These trends together with previous results that show how gas inflows are triggered during galaxy interactions and mergers, generating a variation in the metallicity gradients according to the stage of evolution of the interaction \citep{sillero2017,tissera2022}, among other parameters, highlights the relevance of following the chemical history of baryons along the galaxy formation paths \citep{ma2017}.

Observations of our Galaxy, and others in the nearby universe and at high redshift  contribute to map the chemical abundances of stellar populations and gas in different phases. On the one hand, on our Galaxy, observations of stars in the bulge, disc and stellar halo are providing hints on its assembly histories from the early stages of formation \citep[e.g.][]{helmi2001,sestito2019,carollo2023}. On the other hand, Integrated Field Units surveys such as Califa \citep{sanchez2013Califa} and MaNGA \citep{belfiore2017} provide resolved data on the star formation activity and the metallicity distribution in the ISM and the stellar populations \citep[e.g.][]{baker2023}. The ever growing dataset of high redshift observations, currently led by the James Web Space Telescope, are collecting chemical abundances of baryons up to $z \sim 10 $ \citep{curti2023},  which provide new valuable insights into the star formation activity, metallicity and the relation with their nearby environment  and hence, on the process of galaxy formation \citep{nakayima2023, venturi2024}.

 In this paper, we introduce the Chemo-dynamIcal propertiEs of gaLaxies and the cOsmic web project, dubbed \cielo, which aims at studying the formation and evolution of  galaxies, including both their chemical and dynamical properties.  
 The \cielo~ project aims to perform two LG type haloes and 26 regions mapping different environments excluding galaxy clusters,  which will be, hereafter, referred as the Pehuen\footnote{Pehuen  is the name in Mapundungun of a tree typical of the South of Argentina and Chile.} regions.  Additionally, two LG-like (hereafter, LG) environments were also generated.  This paper presents the project and analysed the main properties of the \cielo~galaxies which cover a stellar mass range of approximately $[10^8, 10^{11}]\rm M_{\sun}$. 
 
 Different aspects of galaxy formation and properties have been already tackled by using  \cielo~galaxies.    \citet{rodriguez2022}  studied the effects of a LG-like environment on infalling disc-like satellites, focusing on the impact of ram pressure, tidal stripping and SN feedback on disc-like satellites as the fall-in into the group potential well.  \citet{cataldi2023} investigated the impact of baryons on the shape of the dark matter haloes of galaxies selected from \cielo~ simulations and Illustris-TNG50. These authors already linked the variation of the shapes to the cosmic web and the direction of infall.  \citet{casanueva2024} developed a semi-analytical model to explore the possible contribution of  Primordial Black Holes to the dark matter component, grafted onto a high resolution environment of ~\cielo~ already run. 
 \citep{GonzalezJara2024} analysed the assembly of stellar haloes of the \cielo~galaxies and their chemical abundances, finding not only that the main satellite contributors shaped the  halo MZR but that smaller satellites left their imprint in the $\alpha$-Fe versus [Fe/H] plane. \citet{Tapia2025} studied the origin of the shape of the gas-phase metallicity gradients. 
 
 In this paper, our aim is to provide more detail information on the ICs and the performance of  the \cielo~galaxies in reproducing the main observed fundamental relations. We particularly explore the mass-size, TFR and  MZR relations at $z=0$. This paper is key to completing the analysis of the already published work as well as those currently in preparation. It will also serve as a reference for future improvements of  or addition to the subgrid physics. Forthcoming papers will analyse in detail different aspects of the formation of these galaxies as the metallicity budget and outflow loading factors, the properties stellar populations in  bulges and discs.

 This paper is organised as follows: Section 2 describes the initial conditions and the version of {\small GADGET-3} used in this study. In Section 3, we characterise the global environment of the Pehuen haloes. Section 4 focuses on the global properties of the \cielo~galaxies and their ability to reproduce key fundamental relations. Finally, the Summary outlines our main findings.

\section{The \cielo~ simulations}

The \cielo~ simulations were carried out using a version of {\small GADGET-3} that incorporates a multiphase model for the gas component, metal-dependent cooling, star formation, and supernova (SN) feedback, as described by \citet{scan05} and \citet{scan06}. These multiphase and SN-feedback models have been successfully used to reproduce the star formation activity of galaxies during both quiescent and starburst phases. They are also capable of driving mass-loaded galactic winds, with an strength corresponding to the depth of the  gravitational potential well of the ga galaxy \citep{scan05,scan06}. This physically motivated SN-feedback scheme avoids the use of arbitrary mass-scale dependent parameters. As a consequence, it is
particularly well-suited for the study of galaxy formation in a
cosmological context. 
The subgrid physics models used to run the current set of ~\cielo~ galaxies has been slightly modified to achieve better results 
regarding the chemical abundances as explained below. 

Our chemical evolution model includes  enrichment by Type II and Type Ia Supernovae (SNII and SNIa, respectively). A Salpeter and Chabrier Initial
Mass Functions are available, with lower and upper mass cut-offs of 0.1
${\rm M_{\odot}}$ and 40 ${\rm M_{\odot}}$, respectively. We follow 12
different chemical isotopes: H, $^4$He, $^{12}$C, $^{16}$O, $^{24}$Mg,
$^{28}$Si, $^{56}$Fe, $^{14}$N, $^{20}$Ne, $^{32}$S, $^{40}$Ca, and
$^{62}$Zn. Initially, baryons are in the form of gas with primordial
abundance X$_{\rm H}=0.76$ and Y$_{\rm He}=0.24$.

SNII are assumed to originate from stars more massive than 8 ${\rm
M_{\odot}}$. Their nucleosynthesis products are derived from the
metal-dependent yields of \citet{WW95}. The lifetimes of SNII progenitors are
estimated according to the metal-and-mass-dependent lifetime-fitting
formulae of \citet{rait1996}. For SNIa, we adopt the W7 model of
\citet{iwamoto1999}, which assumes that SNIa events originate from CO white
dwarf systems in which mass is transferred from the secondary to the
primary star until the Chandrasekhar mass is exceeded, and an explosion
is triggered. For simplicity, we assume that the lifetime of the
progenitor systems are selected at random over the range $[0.7, 1.1]$
Gyr. This choice is supported by the findings of \citet{jimenez2015}, who conducted a detailed analysis comparing the outcomes of this simplified SN Ia lifetime model with the single-degenerate (SD) model \citep{matteucci1986}, revealing very similar trends. To estimate the number of SN Ia events, we adopt an observationally motivated SNII-to-SNIa rate ratio, as described by \citet{mosconi2001}.

The ejection of chemical elements is grafted onto the SN-feedback model,
so that chemical elements are distributed within the cold and hot gas
phases surrounding a given star particle. The fraction of elements that is injected 
into each phase is regulated by a free parameter, $\epsilon_{\rm Z}$, while
the amount of SN energy received by each phase is regulated by $\epsilon_{\rm cold}$.
 Both types of SNe are assumed to inject the same amount of energy into the ISM in a similar fashion. The injection of energy follows two different paths, depending on the thermodynamical properties of the gas, but each phase receives $\epsilon_{\rm cold} = 0.5$. Chemical elements are injected simultaneously with the occurrence of SN events into the hot and cold gas phases, assuming $\epsilon_{\rm Z} = 0.8$. Hence,  80 per cent of the released chemical elements are deposited in the cold gas-phase.  
 The parameters of the subgrid physics used for these runs are taken from the Fenix simulation, which was performed with the same version of {\small GADGET-3}. The Fenix galaxies were also found to reproduce observed relations such as the trend between the specific angular momentum, the stellar mass and the morphology of galaxies \citep{pedrosa2015} as well as the metallicity distribution of the gas and stellar components in the simulated galaxies \citep{tissera2016b,tissera2016}. 

 \begin{figure}
\resizebox{8cm}{!}{\includegraphics{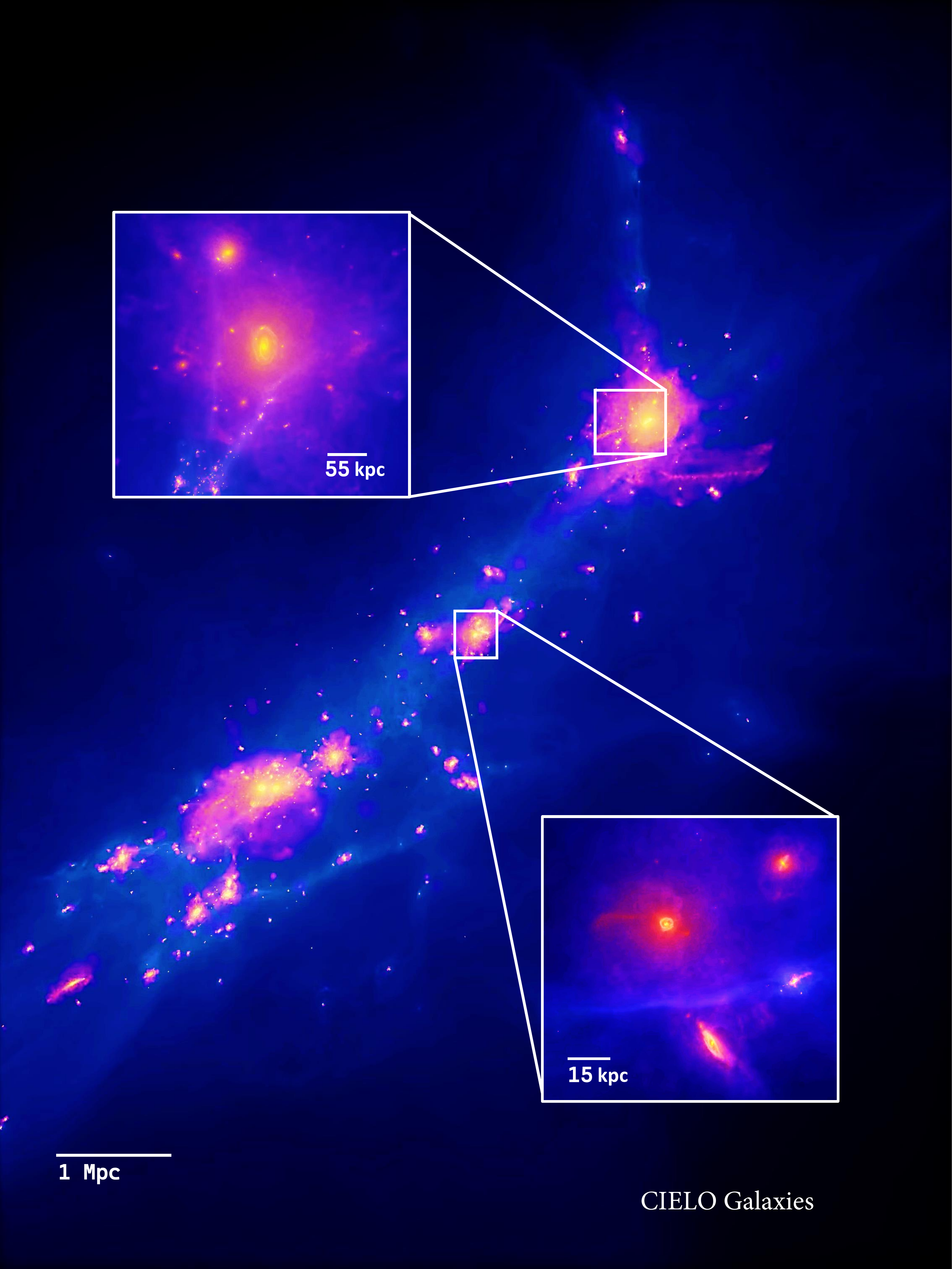}}
\caption{Projected gas and stellar distributions for \cielo~  region corresponding to a filament of region CIELO-P7 (see Table\ref{tab:pehuen_haloes}). The two insets zoom-out two central galaxies, their satellites and closer companions. The image was generated by using Py-SPHViewer \citep{llambay2015}. }
\label{fig:cielogeneral}
\end{figure}

\subsection{Initial conditions}
As mentioned in the introduction, the \cielo~project encompasses two sets of ICs: the LGs and  the Pehuen samples. The former is comprised of two LG-like haloes and the latter includes haloes with a large range of virial masses located at different environments in filaments,  walls and voids (Fig.~\ref{fig:cielogeneral}). None of them include regions as dense as large groups or galaxy clusters. 
The \cielo~simulations assume  a $\Lambda$ Cold Dark Matter universe with $\Omega_0=0.317$, $\Omega_{\Lambda}=  0.683$, $\Omega_B=0.049,$ h$=0.6711$, $n =0.962$ \citep{Planck2013}.

The ICs were generated at two resolution levels, using  dark matter particles with masses,  $\rm m_{\rm dm}= 1.36\times 10^5 \Msun h^{-1}$ for L12 level and $\rm m_{\rm dm}=1.28\times 10^6 \Msun h^{-1}$ for L11 level (high resolution and intermediate resolution, respectively).
The initial gas masses are $\rm m_{\rm gas}= 2.1\times 10^4 \Msun h^{-1}$  and $\rm m_{\rm gas}=2.0\times 10^5 \Msun h^{-1}$
for L12 and L11 levels, respectively.
Simulations at L11 level of resolutions have gravitational softening of $\epsilon_{\rm g}= 400$, 400, and 800 physical pc for the gas, stellar and dark matter  particles, respectively, whereas L12 runs have $\epsilon_g =$  250, 250 and 500 pc, respectively.

There are 128 snapshots distributed between $z\sim 60$ and $z=0$ available to study the evolution of the properties of the structure and galaxies with time. These simulations provide us with a suitable cadence ($\sim 0.17$ Gyr)to follow the impact of different physical processes as satellite galaxies fall into their main haloes, the evolution of the star formation, gas infall and chemical enrichment, among other properties and relations across time. 

In the next sections, we describe the main characteristics of galaxies in both samples. This is a long-term project,  hence, in this paper we present the whole set of selected ICs. The main characteristics of both sets of simulations  are summarised in Table\ref{table1} and Table~\ref{tab:pehuen_haloes}.
In the latter, we include the classification of the local and global environment explained in the following section.

\subsubsection{The Local Group analogues}
The LG analogues are taken from a dark matter-only run of a cosmological periodic cubic box of side length $L = 149$ Mpc  with the adopted cosmology.  The MUSIC code \citep{HahnAbel2011}, which computes multi-scale cosmological initial conditions under different approximations and transfer functions, was applied to extract the selected regions and increase the numerical resolution. Later on, baryons were added to the ICs.

We used the Rockstar halo finder \citep{behroozi2013} to identify and characterise
haloes in the reference simulation.
A first set of 20 LG-type systems was selected from which two LG analogues were resimulated.
Hereafter, they will be named LG1 and LG2. 
The two main haloes within the LG1  have a relative velocity of   $\Delta V = 165.4$\ \kms\
and a physical separation of   $\Delta r = 0.69 \ \rm Mpc$ at $z=0$, while for LG2, $\Delta V=  116.1$ \kms \ and $\Delta r = 0.79 \ \rm Mpc$.
In Table~\ref{table1} we summarise the  main properties of the LG analogues.

\begin{table}[]
\caption{Main characteristics of the Local Group analogues. }

\begin{tabular}{l c c c c c c}
\\ \hline \hline 
 LG  &  ID & \mvir &   x     &     y &         z &           $\Delta V$ \\
 & & $10^{12}\Msun$ & kpc  & kpc  & kpc & \kms\\
 \hline 
 LG1 &0 & 1.9    &  51.5  &     49.6 &     44.1 &     220.0\\
 &1 & 1.2    &  51.8  &    50.0  &    43.9   &   188.5\\ 
 LG2&0 & 7.7    &  48.2  &    48.6    &  51.9  &    162.4 \\
 &1 & 7.4    &  48.4  &    48.4    &  52.3 &    160.6 \\
\end{tabular}
\label{table1}
\tablefoot{From left to right:  the halo ID , \mvir~ and \rvir,  the centre of masses of the two main galaxies and their relative velocities.}
\end{table}

\subsubsection{The Pehuen haloes}

The Pehuen haloes correspond to  24 zoom-in initial conditions  generated with the code MUSIC. In this case, the reference simulation is a dark matter only cosmological periodic cubic box of side length $L = 50$ Mpc $h^{-1}$, consistent with the same cosmology as the LG sample. For consistency, we used the Rockstar halo finder for the identification and characterisation of haloes in this reference simulation. Halo candidates for resimulation satisfied the following criteria to ensure that they are the dominant halo within their immediate environment.

\renewcommand{\labelitemi}{$\bullet$}
\begin{itemize}
 \item The halo virial mass is in the range [ $10^{11}$ - $10^{13} \Msun$]  .
 \item The halo virial mass is at least four times the mass of the most massive neighbour
 inside the  virial radius of a halo.
 \item There is no neighbouring halo more massive than $1.5 \times 10^{13} \Msun$  
 within  $1.5$ Mpc  radius.
\end{itemize}

Potential candidates are then sorted into six mass bins and five environmental bins.
Environment is characterised using the $D_{5,f}$ parameter \citep{hass2012}, an environmental parameter which is insensitive to the mass of the central halo. $D_{5,f}$ is dimensionless quantity defined as the three-dimensional distance to the $5^{\rm th}$ neighbour with a viral mass that is at least $f$ times that of the halo under consideration, divided by the virial radius of the $5^{\rm th}$ nearest neighbour. Throughout this work, we adopt a value of $f=1$.
The final targets were, then, randomly selected to cover the already indicated mass range and
environmental bins. In this paper, we present the first set of Pehuen haloes as shown in Table~\ref{tab:pehuen_haloes}.

\begin{figure}
\includegraphics[width=0.45\textwidth]{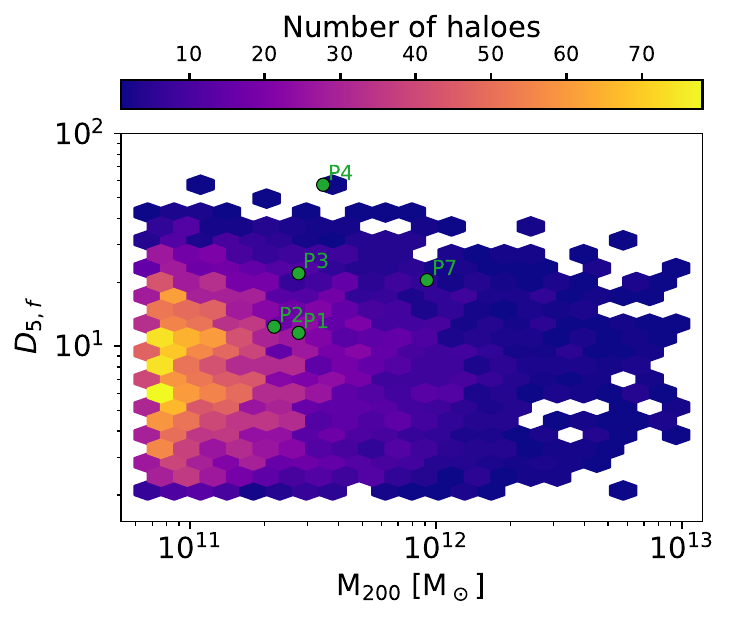}
\caption{$D_{5, f}$ parameter as a function of virial mass for all haloes with virial mass larger than $10^{11}$ M$_\odot$ h$^{-1}$. The green labels indicate the position of the haloes that are at the centre of the zoom-in regions selected for the simulated runs  presented in this paper. }
\label{fig:dnf_vs_mvir}
\end{figure}

Figure \ref{fig:dnf_vs_mvir} shows the $D_{5, f}$ parameter as a function of virial mass for all haloes with virial mass larger than $1.5 \times 10^{11}$ M$_\odot$  (the labels indicate the position of the regions that have already been run). 
The selected haloes cover the full range of virial masses and environments almost uniformly, avoiding too dense regions. This is a characteristic of the \cielo~ simulations. The current set of zoom-in regions is
centre at a similar mass halo but with different local environments (P1, P2 and P3). To have a larger diversity we also included a smaller halo in a denser local region (P4) and a more massive central halo in an intermediate local region (P7). The latest was run at the two level of resolution (L11 and L12) to check the effects of numerical resolution.

\begin{table*}[]
\caption{Main characteristics of the target Pehuen haloes. }  
\label{tab:pehuen_haloes}
\begin{tabular}{ccccccccccc}
\hline\hline
Pehuen  &  ID & \mvir & \rvir & $D_{5, f}$&$d_{\rm node}$ & $d_{\rm  min}$ & $d_{\rm  sadd1}$& $d_{\rm sadd2}$ & $d_{\rm  fil}$& E$_{\rm nv}$\\
& & $10^{11}$$ \Msun $ & $\mathrm{kpc}$  &   &  $\mathrm{Mpc}$ &  $\mathrm{Mpc}$  &  $\mathrm{Mpc}$  &  $\mathrm{Mpc}$ &  $\mathrm{Mpc}$  \\
\hline 
P1& {33306}	& 4.1 & 196.7  &  17.3 &  3.4  & 2.5  & 7.5	& 3.7	& 3.4 & void\\
P2& 206641	& 3.3 & 181.8  &  18.5 &  9.2  & 5.2  &  5.8	& 4.8	& 4.2 & filam\\
P3& 229533	& 4.1 & 196.7  &  32.8  & 8.9	& 7.6  & 3.1 	& 2.7   & 1.0&  filam\\
P4& 124429	& 5.7 & 211.6  &  86.7  & 8.8 	& 4.5  & 3.6    & 5.5	& 4.5 & wall\\
P7& {147062}& 13.7 & 293.5  & 30.5  & 9.8	& 13.3  & 6.9  	& 5.8	& 5.1 &filam \\
\end{tabular}
\tablefoot{From left to right:  the target halo ID and its \mvir~ and \rvir~assigned by the Rockstar halo finder,  the $D_{5, f}$ local environmental parameter by \citet{hass2012}, the distribution of distances to the closest of each of the DisPerSE critical points: nodes, voids, 1D-saddles, 2-d saddle and filaments, and the global environment classification \citep{Sousbie2013}.}
\end{table*}

The high-resolution regions are determined within $x$ times the virial radius of the central haloes, followed by defining the minimum ellipsoid that encloses these particles at $z=60$ \citep{onorbe2015}.The $x$ factor is determined individually for each halo. DMo simulations for different values of $x$ factors were run. Then, we selected the smallest value of $x$ for which no low-resolution particles (boundary particles) are present within the virial radius of the target halo at $z=0$. In general, $x$ ranges from 2 to 8\rvir, depending on the halo environment.

\subsection{The global environment of the Pehuen  haloes}
\label{environ}

This subsection focuses on quantifying the characteristics of the cosmic web environments for the Pehuen haloes at $z=0$. We investigated the outcomes of the DisPerSE \citep{Sousbie2013} method, which we employed to segment the cosmic web into its components, using the initial $75$ Mpc  box from which the Pehuen haloes were extracted. The detection of filaments and other elements of the cosmic web relies on a two-step process. First, the matter density field is measured, followed by the application of discrete Morse theory \citep{Morse1934}, which is then used to detect filaments.

\begin{figure}
\includegraphics[width=0.45\textwidth]{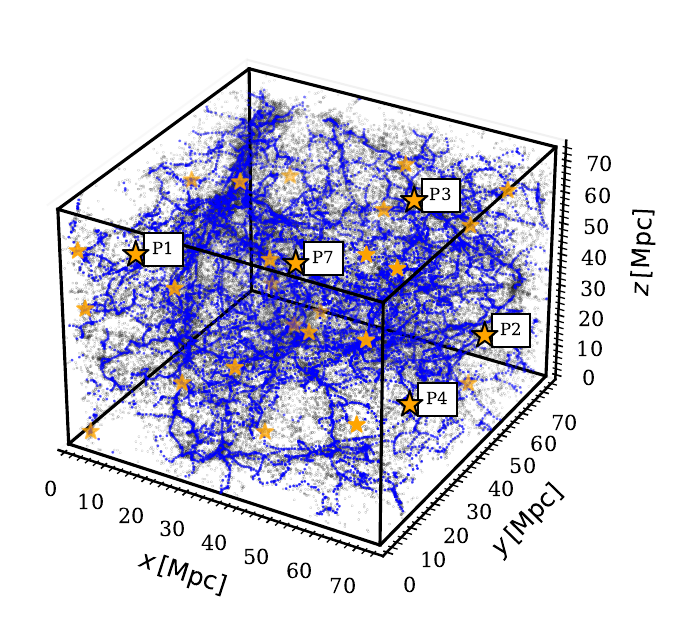}
\caption{Cosmic filaments (blue lines) detected by DisPerSE method for the full dark matter distribution at $z \sim 0$. The position of the whole Pehuen regions and those presented in this paper are highlighted (yellow stars and labels, respectively).}
\label{fig:box_filaments}
\end{figure}

\begin{figure}
\centering
\includegraphics[width=0.45\textwidth]{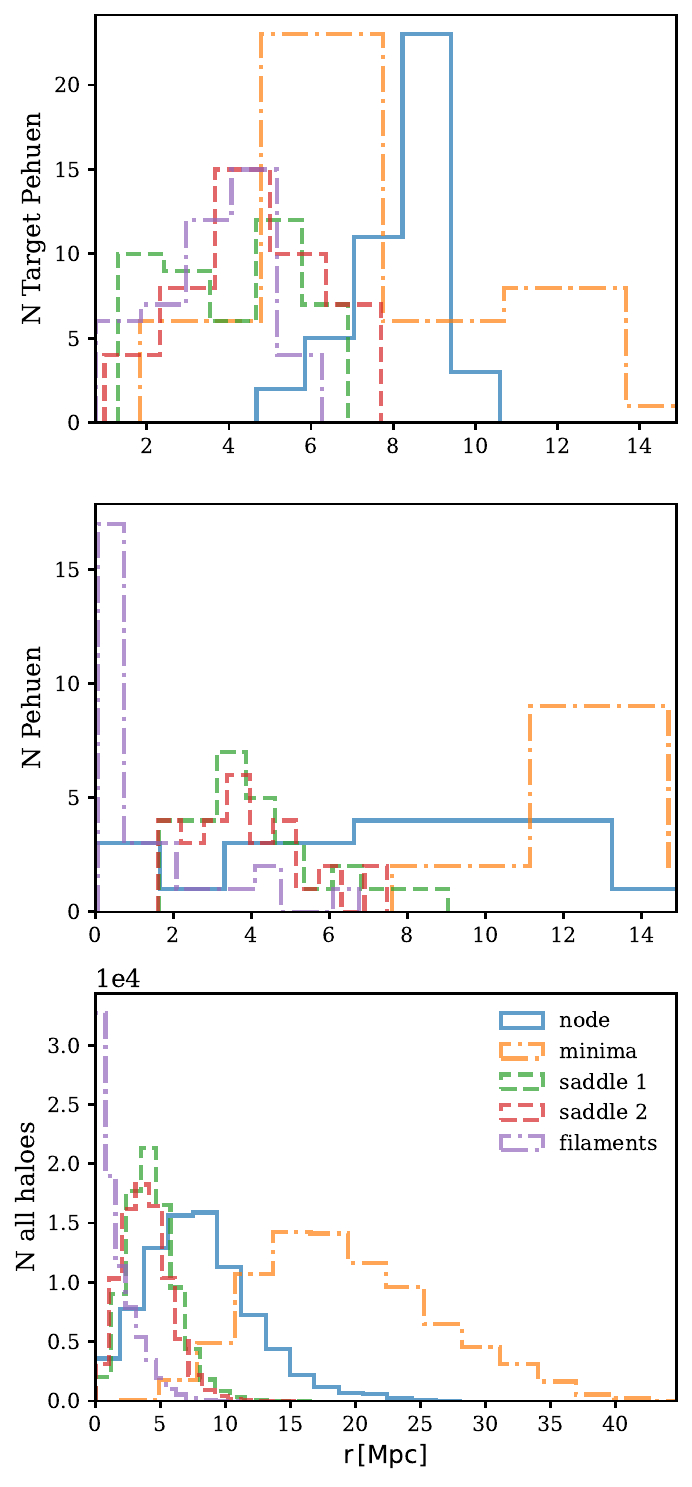}
\caption{Environment distribution quantified to the closest distance to each cosmic-web structure, according to DisPerSe. The distributions are displayed for the zoom-in regions studied in this paper (see Table~\ref{tab:pehuen_haloes}; top panel), for the whole target sample of Pehuen haloes (middle panel) and for  haloes in the whole initial box (lower panel).}
\label{fig:hist_fila}
\end{figure}

We reconstructed the halo-number density field  using the Delaunay tessellation field estimator \citep[DTFE,][]{Schaap2000, vanderWeygaert2009}. This method uses Delaunay tessellation to cover the space with tetrahedrons, using the halo positions as vertexes. The density field can be recursively smoothed by averaging the value at each halo position with the density field values of the galaxies that are directly connected to it through an edge of the tessellation.

The DisPerSE method then proceeds with the application of the discrete Morse theory to the measured density field. DisPerSE is a multiscale structure identifier that detects persistent topological features such as peaks, voids, walls and, especially, filamentary structures. For the discrete set of points used to estimate the density field, we chose haloes from the original dark matter-only, DMo, box with a minimum mass of $10^{11} M_{\odot}$, which is a strong bound for the mass of halos that could form galaxies, in a hydrodynamical scenario.

We employ a value of $\sigma =3$ for the ‘persistence’ parameter, which represents the robustness of the critical point determination. This value of persistence for the filamentary detection gives the best coincidence of positions of the maximum density critical points for the over-dense regions \citep[see e.g,][for the implementation in numerical simulations]{Galarraga2020}. For the procedure and technical aspects, we refer the reader to the DisPerSE webpage. \footnote{\url{http://www2.iap.fr/users/sousbie/web/html/indexd41d.html?}}

The cosmic web can be described using a mathematical framework known as the Morse complex, which consists of a set of manifolds that, in the cosmological context, correspond to the structures mentioned above. In Fig.~\ref{fig:box_filaments} we display the filamentary structure of the simulated volume from which the Pehuen sample is drawn. The labels denote the zoom-in regions presented and analysed in this paper.

The critical points of the density field, which correspond to the locations where the gradient of the density field vanishes, can be derived (i.e. the maxima, minima or saddle points of the density field). Nodes define the global maxima, while voids are associated with minima points. 1-D and 2-D saddle points are local density minima bounded to structures, such as walls or filaments, respectively. The 1-D saddle points are inside walls, whereas the 2-D saddle points are located within filaments. Finally, filaments are defined as field lines of constant gradient connecting critical points, composed of short segments of which the positions of the extrema are given. The length of the segments is related to the typical length of the edges of the tessellation.

We proceed to classify each halo according to their distance to the nearest substructure (node, minima, saddle-1, saddle-2 and filaments). While this classification does not directly represent the local environment of a given galaxy or halo,  it indicates their proximity to specific cosmic web features described above. Low distances to nodes correspond to high-density regions, whereas low values of distances to filaments trace the filamentary network. Low values of distances to minima mark the low-density regions and saddle points are found mostly inside walls and filaments. In Table \ref{tab:pehuen_haloes} we present for the first set of Pehuen haloes their distance to the nearest DisPerSe critical point. Pehuen haloes were chosen precisely to have a diversity of environments, which is reflected in the different minimum values of each halo with the different critical points of the density field, whereas in Fig.~\ref{fig:hist_fila} we show the distributions of distances to the before-mentioned analysed Pehuen haloes (top panel), the whole selection (middle panel) and for all haloes in the simulated volume from which the Pehuen sample is obtained (bottom panel). 

Pehuen haloes show distances much closer to the critical points characteristic of the cosmic web. This is expected, as the Pehuen haloes were specifically selected for their location in diverse environments. For the overall results of the entire simulated DMo box, the haloes follow distributions already reported in previous works for simulations and for galaxy surveys \citep[see for example,][]{Malavasi2020,Montero-Dorta2024}. Finally, we used the classification according to their minimum distances from critical points to characterize the environment of each re-simulated region (see Table~\ref{tab:pehuen_haloes}). 
We stress the fact that the  $D_{5,f}$, quantifies the local environment while the DisPerSe classification is based on the large-scale distribution. In a future work, we will explore the dependence of galaxy properties on both parameters.

\subsection{The \cielo~ galaxies}
All \cielo~simulations are analysed by using the same method.
Haloes are identified at their virial radius, \rvir, by using a friends-of-friends  algorithm \citep[FoF,][]{davis1985}. The substructures are selected by using the {\sc SUBFIND} algorithm \citep{springel2001a, dolag2009}. The merger trees were built by using the {\sc AMIGA} algorithm \citep{amiga}.

\subsubsection{Mocking SDSS images}
We generated synthetic optical images for the \cielo~ galaxies using the 3D radiative transfer code \texttt{SKIRT} \citep{baes2011,camps2015}, which simulates the physical processes that radiation undergoes while propagating through a medium. In this work, we used the latest version, \texttt{SKIRT 9} \citep{camps2020}. For each galaxy in our sample, we selected stellar particles within a 50 kpc aperture to serve as radiation sources. Each particle was assigned a stellar template from the \citet{BC2003} library, based on its age and metallicity in the simulation snapshot, assuming a \citet{chabrier2003} IMF.

The diffuse dust was modelled based on the properties and spatial distribution of gas in the simulation, assuming that a portion of the ISM gas contains dust. We applied the recipe from \citet{torrey2012}, which selects dust-carrying gas particles based on their position in the temperature-density phase space, as follows $\log(\rm T/K) < 6 + 0.25 \log_{10}(\rho_{\rm gas} / (0.45 \times 10^{10}) \  M_{\odot} \ \rm kpc^{-3})$. Dust was then allocated to these particles using a fixed dust-to-metal ratio, defined as $\rho_{\rm dust} = f_{\rm dust}Z\rho_{\rm gas}$, where $Z\rho_{\rm gas}$ is taken from the simulations. The calibration from \citet{kapoor2021}  was used for $f_{\rm dust}$. For the dust properites, we adopted the THEMIS dust model \citep{jones2017}.

\begin{figure*}
    \centering
    \includegraphics[width=1\textwidth]{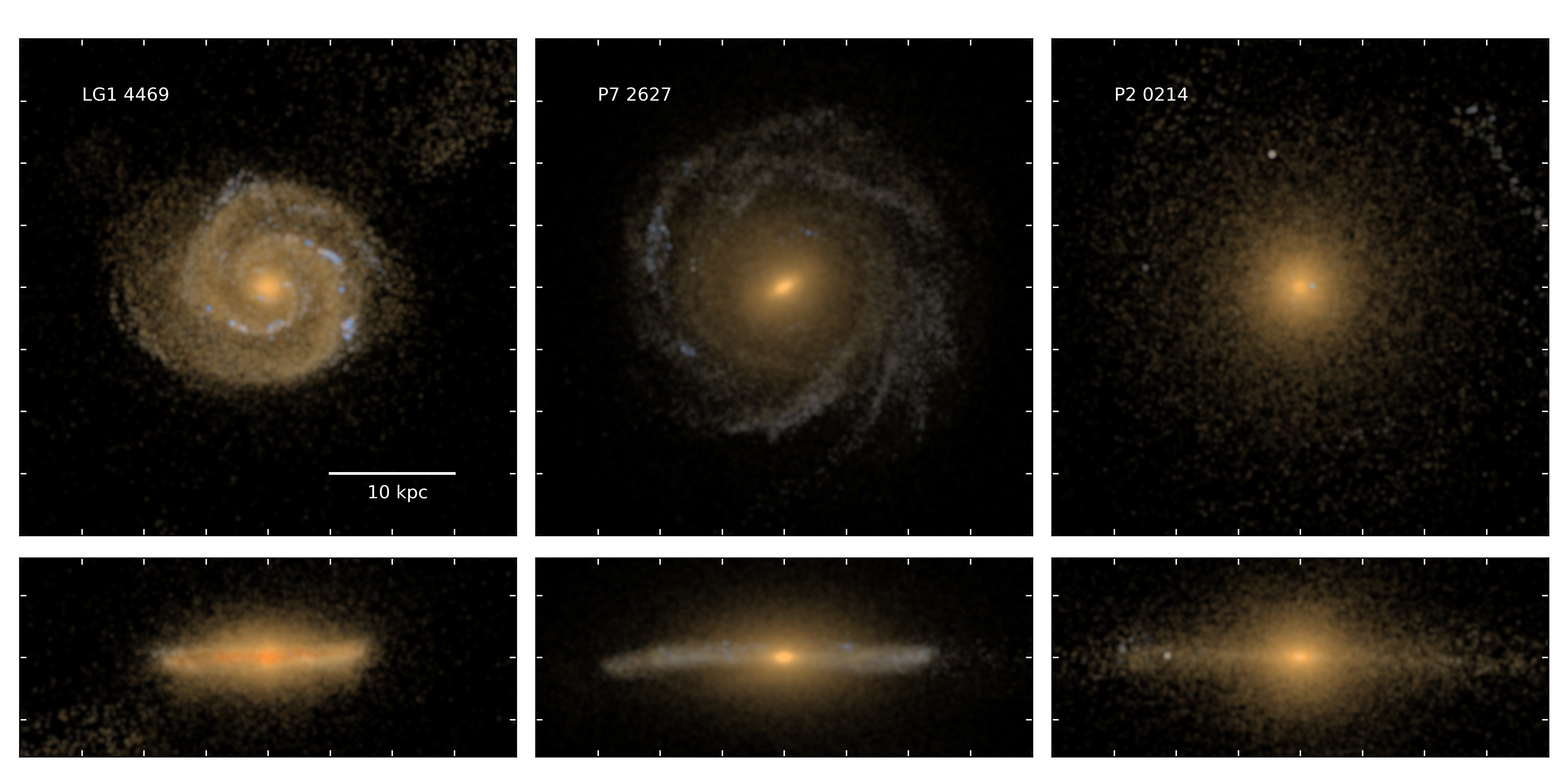}
    \caption{Face-on (top row) and edge-on (bottom row) views of selected \cielo~ galaxies. Images are colour-composites using synthetic SDSS i, r, and g observations generated using the 3D radiative transfer code SKIRT, and following the methods and calibrations from \citet{kapoor2021}. Galaxies are placed at a distance of 20 Mpc and the field of view of all images is $40\times40$ kpc at a pixel resolution of 100 pc on one side.}
    \label{fig:galaxies_skirt}
\end{figure*}

Finally, galaxies were placed at a distance of 20 Mpc from the synthetic instruments and broad-band images were generated at face-on and edge-on viewing angles. These images cover a set of broad-band filters, including GALEX FUV and NUV, SDSS u, g, r, i, and z, as well as Herschel PACS 100, 160 $\mu$m, and SPIRE 250, 350, and 500 $\mu$m. All images were produced with a field of view (FOV) of $40 \times 40$ kpc$^2$ and a spatial resolution of 100 pc pixel$^{-1}$. Composite colour images were created using SDSS i, r, and g filters, as shown in Fig.~\ref{fig:galaxies_skirt}. In this figure we see a set of three spiral galaxies as examples, each of them with different types of arms and relevance of the bulge component. These three galaxies are taken as examples hereafter. The images for all the 54 central \cielo~ galaxies are available as part of the CIELO database\citep{GonzalezJara2024}.

\subsubsection{Dynamical decomposition and galaxy morphology}
\label{sec:decom}

In order to analyse the global properties of the \cielo~galaxies, we applied the dynamical decomposition into disc, bulge, and stellar haloes described   by  \citet{tissera2012}, the so-called AM-E method. The method is based on a combination of the angular momentum content of the particles and their binding energy, and avoid the use of strict radial thresholds \citep[see also][]{pedrosa2015}. 
According to the AM-E method, the galactic components were decomposed dynamically based on the circularity parameter, $\epsilon$\footnote{$\epsilon = \frac{J_z}{J_{\rm z, max(E)}}$, 
where $J_z$ is the angular momentum component perpendicular to the disc plane and $J_{\rm z,max(E)}$ is the maximum $J_{\rm z}$ over all particles of given total energy (E).}.

We also estimated the optical radius, $R_{\rm opt}$, as the one that enclosed 80 per cent of the stellar mass identified by the SUBFIND algorithm. The stellar half-mass radius, $R_{\rm hm}$, was calculated as the one that encloses half of the stellar mass. All parameters such as the star formation rate, SFR, the stellar mass, \mstar,  were estimated within the $R_{\rm opt}$ or $R_{\rm hm}$. 

The bulge component is defined as all stellar particles more bounded than the minimum energy, E$_{\rm bin}$), at $r \sim 0.50 \times $\ropt. The disc component is defined as the stellar particles with $\epsilon \geq 0.5$,  E $\leq \rm E_{\rm bin}$ and $r \leq 1.5$\ropt. 
 Stellar particles that do not belong to the bulge and disc components are considered to be part of the stellar haloes \citep{GonzalezJara2024}. 
Therefore, the method adapts to the characteristics of the mass distribution of each galaxy. Although there is some arbitrariness in these criteria, the main features of the components change only slowly with the reference energies.  This decomposition allows us to estimate  the bulge-to-total (B/T) stellar mass ratio and use it for morphological classification. We note that the bar components are not individualised by this method, so bulges might be overestimated in some cases.
For some galaxies, it is also possible to identify a well-defined counter-rotating disc composed of stellar particles with $\epsilon \leq -0.5$, which are made of accreted stars \citep[e.g.][]{carollo2023}.

\begin{figure*}
\includegraphics[width=1\textwidth]{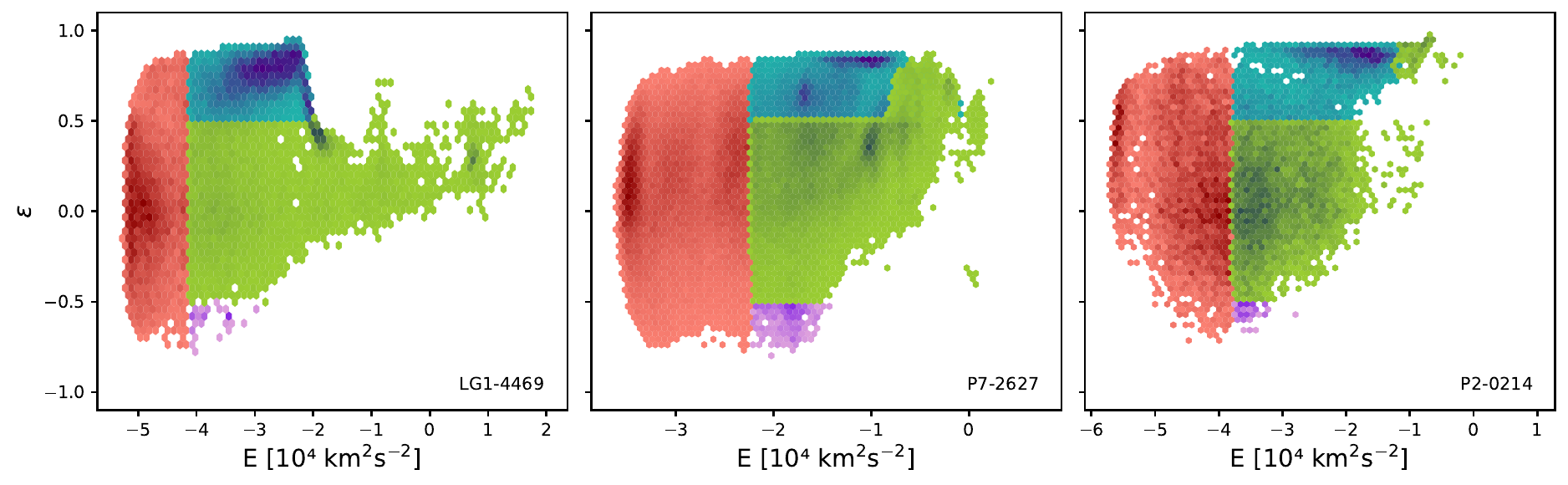}
\caption{Circularity, $\epsilon$, as a function of binding energy, E$_{\rm bin}$, for the three galaxies taken as examples and displayed in Fig.~\ref{fig:galaxies_skirt}. 
The different colours denote the stellar particles classified as bulge (blue dots), disc (orange dots), counter-rotating disc (magenta dots), and stellar halo (green dots).
}
\label{fig:circularities}
\end{figure*}

 For visualization purposes,  Fig.~\ref{fig:circularities} shows the distribution of $\epsilon$  as function of  E$_{\rm bin}$ for the galaxy sample  shown in Fig.~\ref{fig:galaxies_skirt} (only hexbins with more than 10 stellar particles are shown). This figure highlights star particles, which have been classified as part the bulge, the disc, counter-rotating disc, and the stellar halo  of the three galaxies. The selected examples show a diversity of distributions with different relative importance of the different components.  In particular, the B/T $= 0.40, 0.65, 0.66$ for LG1-4469, P7-2627, and P2-0214, respectively.
In the following sections, we analyse the global properties, and the fundamental relations followed by the \cielo~galaxies, using the morphology as a secondary parameter.

\section{Global properties of the \cielo~ galaxies}
In this section we will analyse the global properties of  central \cielo~ galaxies with stellar masses \mstar$ \geq 10^8 \Msun$. They are all measured within \ropt~ as defined in the previous section. 

\begin{figure}
\includegraphics[width=0.45\textwidth]{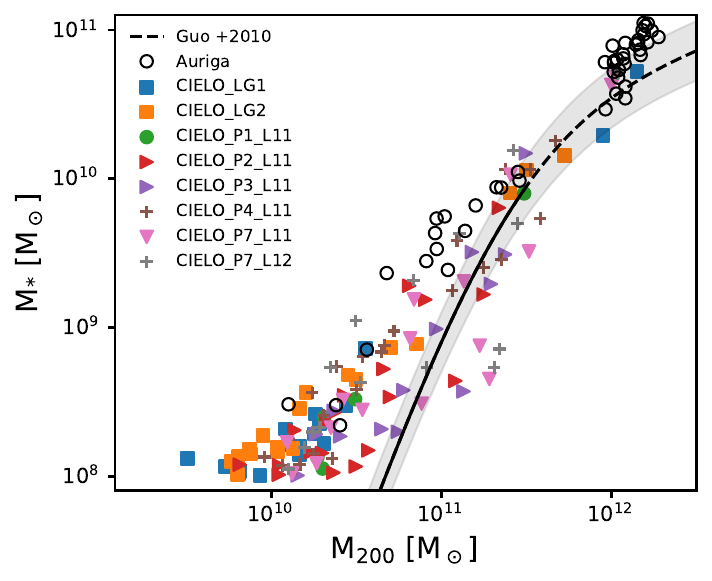}
\caption{Stellar mass as a function of \mvir~ for the central galaxies of haloes with \mstar$\geq 10^{8} M_{\sun}$ for the all analysed simulations (see labels in the inset). For comparison, the relation proposed by \citet{guo2010} (dashed line and the extrapolation for halo masses lower than $10^{11.5}$ indicated by the solid line) and trend for the L4 Auriga galaxies \citep{grand2024} are also included (open, black circles). 
}
\label{mstarmhalo}
\end{figure}

In Fig.~\ref{mstarmhalo} we show the \mstar versus \mvir~relation. As can be seen from this figure,  \cielo~galaxies contained a stellar mass for a given dark matter halo in good agreement with the relation proposed by \citet{guo2010}  for galaxies with \mstar~$\geq 10^{9}\rm M_{\sun}$. These are our main target galaxies. For galaxies with \mstar$=[10^{8}, 10^{9}]\rm M_{\sun}$, there is an excess with respect to the expected observationally motivated relation. However, the trends agree with the extrapolation and results from other simulations \citep[e.g.][]{sales2022,grand2024}. 
At the low mass end, there is a larger scatter, or a larger diversity of \mstar~ at a given \mvir~ as reported by other cosmological simulations \citep{onorbe2015,maccio2017,sales2022}.  We also estimate this relation for galaxies in CIELO-P7-L11 and CIELO-P7-L12. They are both included in this figure. However, to make the comparison clearer, Section \ref{appendix} presents this relation for only these two runs. Both are in good agreement with observations as displayed in Fig.~\ref{fig:comparison}.  
We also analysed a possible dependence of this relation on environment by using the classification summarised in Table~\ref{tab:pehuen_haloes}. 
However, from this set, it is not yet possible to assess reliably the dependence on environment because of the low statistical number, and hence we delay a comprehensive analysis of the global environment for a future work.

\begin{figure}
\includegraphics[width=0.45\textwidth]{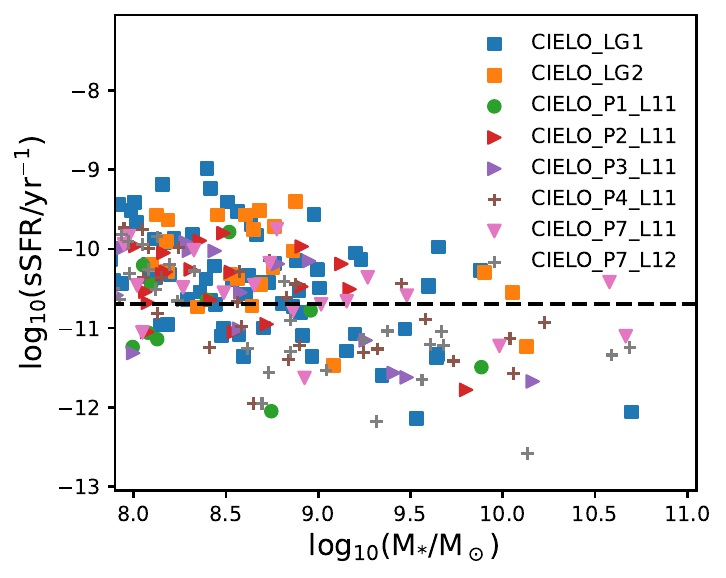}
\caption{sSFR as a function of  stellar mass for galaxies with  $M_{\star} \geq 10^{8} M_{\sun}$. To increase the statistics satellites galaxies  within the same mass range are also included. The dashed line depicts the adopted limit for star-forming galaxies  of \logten$\rm(sSFR yr^{-1}) = 10^{-10.8}$. Galaxies tend to have lower sSFR at the higher mass end (see also Fig.~\ref{ssfr_hist}). 
}
\label{ssfr}
\end{figure}

\begin{figure}
\includegraphics[width=0.45\textwidth]{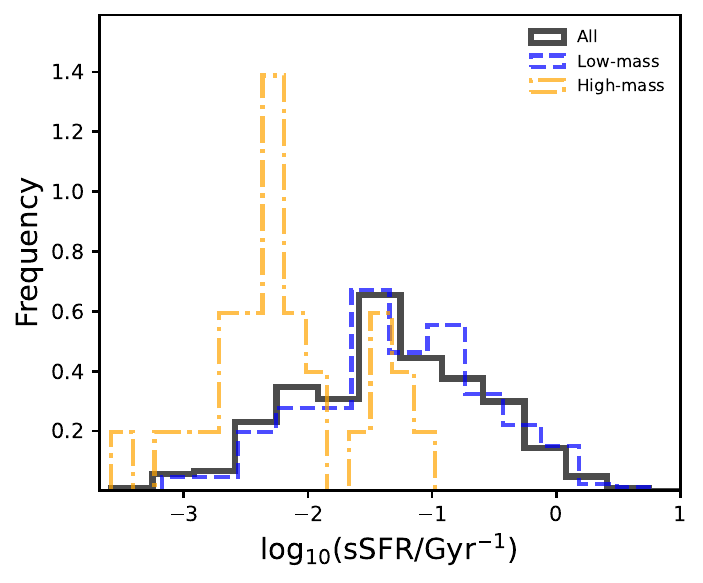}
\caption{Distribution of sSFR for \cielo~ galaxies with  \mstar~$ \geq 10^{8}\rm M_{\sun}$ (black solid lines). The distribution normalised by unit area shows hints of bimodality.
The corresponding distributions for galaxies with  \mstar~$[10^{8}, 10^{9.5}]\rm M_{\sun}$ (dashed gold lines) and with  \mstar~$[10^{9}, 10^{11}]\rm M_{\sun}$ (blue lines) are also displayed to highlight the bimodality. 
}
\label{ssfr_hist}
\end{figure}

In Fig.~\ref{ssfr} we display the specific star formation rate, defined as sSFR~$=\rm SFR/$\mstar, as a function of stellar mass. The dashed line depicts  \logten~(sSFR/ yr$^{-1}) = -10.8$, taken as a reference to separate star-forming galaxies from quiescent ones in the local universe \citep{dave2019}. The \cielo~ galaxies reproduce the observed anticorrelation between sSFR and \mstar. 
However,  considering that we have a small sample from different low-density regions instead of a large cosmological volume, to better assess the distribution of sSFR in Fig.~\ref{ssfr_hist},  we show the probability distribution of sSFR for  massive and low-mass galaxies, separately, adopting \mstar$=10^{9.5}\Msun$ as a threshold.  As shown, the distribution is bimodal and generally consistent with observations. However, we note that the peak of the distribution of the star-forming galaxies is slightly shifted toward lower sSFR compared to the results shown in \citet{dave2019}. Nonetheless, given the low-number statistics, the \cielo~galaxies overall appear to reasonably reproduce the level of star formation activity per unit stellar mass (solid black line). The distributions  for low-mass (dashed blue lines) and for massive (dotted-dashed gold lines) star-forming systems are also consistent with observations, considering that our sample is still small \citep[see also][]{salim2018}. 

\subsection{Fundamental relations of the \cielo~ galaxies}

In the following section, we analyse the fundamental relations determined by the \cielo~galaxies and compare them with observations. We only consider in our analysis central galaxies resolved with more than 1000 stellar particles in the L11 runs and more than 10000 in the L12 runs. These implies a lower stellar mass limit of about $2 \times 10^8\Msun$. The dynamical decomposition described in the previous section is applied to the whole sample of 54 systems.

\begin{figure}
\includegraphics[width=0.45\textwidth]{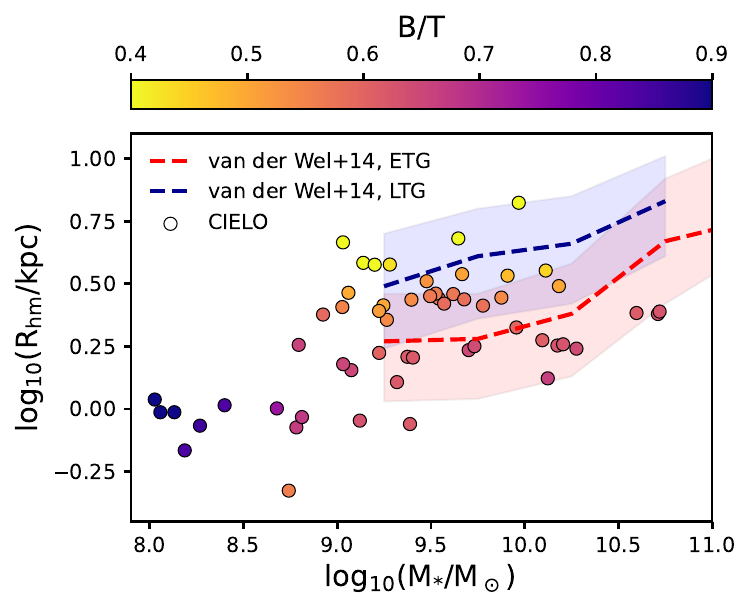}
\caption{Mass-size relation for the \cielo~ galaxies defined by the stellar half-mass radius, \reff,  as a function of the B/T ratio. More disc-dominated galaxies have larger sizes at a given stellar mass as expected.  The median observed trends for late-type  (LTG;  dashed blue lines) and early-type galaxies (ETG; dashed red lines) reported by \citet{vanderwel2014} are shown (shaded regions are determined by the 25-75$^{\rm th}$ percentiles (LOESS). 
}
\label{fig:half_mass}
\end{figure}

Figure~\ref{fig:half_mass} displays the mass-size relation as a function of the bulge-to-total stellar mass ratio, B/T. For this analysis we defined size as the radius that enclosed half the stellar mass of the galaxy within \ropt. We used the 2D locally weighted regression method to generate smoothed distributions \citep[LOESS;][]{Cappellari2013}. Galaxies with more prominent bulges tend to have lower masses or to have smaller sizes at a given stellar mass. We note that the \cielo~sample does not include massive dispersion-dominated galaxies by construction (see Table\ref{table1} and Table\ref{tab:pehuen_haloes}).
As shown in this figure, at a given stellar mass, more disc-dominated systems are more extended as expected \citep{vanderwel2014}. Additionally, we found a clear anti-correlation signal between \reff and B/T with Spearman coefficients of  $r =-0.62$ and $p < 0.001 0$.

A basic relation that has to be satisfied by simulated disc galaxies is the TFR \citep{tully1977, dutton2011}. Hence, we estimate the  TFR  using the maximum rotational velocity, V$_{\rm max}$, \citep[e.g][]{lapi2018}. 
 For each galaxy, we constructed the rotational curves of the gas component by using the rotation velocities after aligning the total angular momentum along the z axis (as explained  in Section~\ref{sec:decom}). Then, V$_{\rm max}$  is defined as the maximum value reached by the rotational curves (Miranda et al. in prep.). 
 
\begin{figure}
\includegraphics[width=0.45\textwidth]{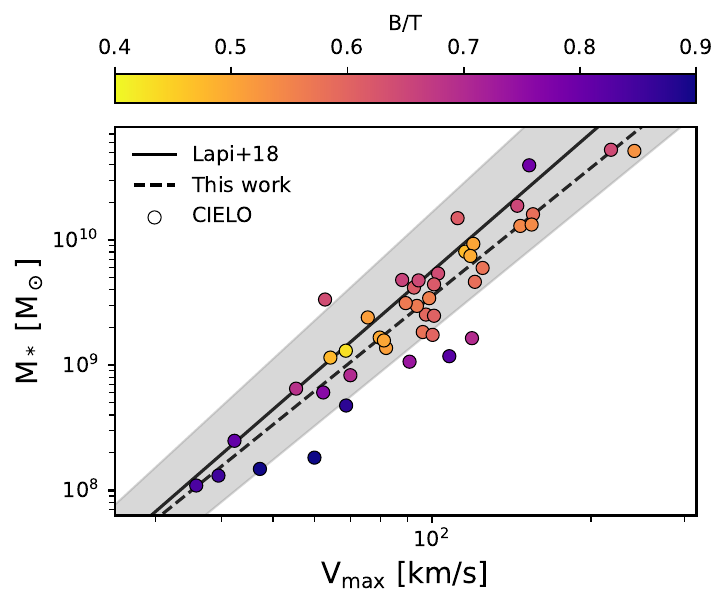}
\caption{Stellar  TFR for the \cielo~galaxies (circles).  The linear regression fit  is also displayed (dashed line).  For comparison  observational results for galaxies in the Local Universe by \citet{lapi2018} has been included (solid line). The shaded region represents the observed standard deviation. 
}
\label{BTFR}
\end{figure}

 In Fig.~\ref{BTFR}, the stellar TFR is shown for our simulated sample, with colours representing the LOESS-smoothed dependence on B/T. For comparison,  the observed relation reported by \citet{lapi2018} for galaxies in nearby groups and diffuse clusters, and field galaxies is shown. These authors also assumed a Chabrier IMF.   
As can be seen in Fig.~\ref{BTFR}, \cielo~galaxies follow the observed trend quite well.  The simulated galaxies show slightly less mass at a given V$_{\rm max}$. However, most of them are within the standard deviation. We note that the \cielo~ galaxies are tracing systems in low-density environments, while the observed relation also includes galaxies in clusters. We fitted a linear regression to the \cielo~galaxies that yield: \logten(\mstar$/\Msun$)$= 3.41\pm 0.22 \times {\rm log}(V_{\rm max}/{\rm km s^{-1}}) + 2.73 \pm 0.44$. The observational relation reported by \citet{lapi2018} is  \logten(\mstar$/\Msun$) $=  3.67\pm 0.23 \times {\rm log}(V_{\rm max}/{\rm km s^{-1}}) + 2.41 \pm 0.10$. We note that the simulated stellar mass are directly extracted from the simulation, while observations have to infer it from stellar population and dynamical modelling. Additionally, instrumental effects (beam smearing) and the presence of dust have been shown to flat rotation curves, leading to an underestimation of the  rotation velocities \citep[e.g.][]{deugenio2013, vandesande2018, barrientosacevedo2023}. This could introduce differences; however, the observed and simulated values remain within the error bars.

\subsection{The mass-metallicity relation}

One of the main objectives of the \cielo~Project is to explore the relationship between chemical abundances and the dynamical properties of stellar populations and gas in galaxies. A key relation for galaxies is the MZR, which we estimate and analyse for both star-forming gas and stellar populations. For both relations, we present the LOESS-smoothed dependence on B/T.

\begin{figure}
\centering
\includegraphics[width=0.45\textwidth]{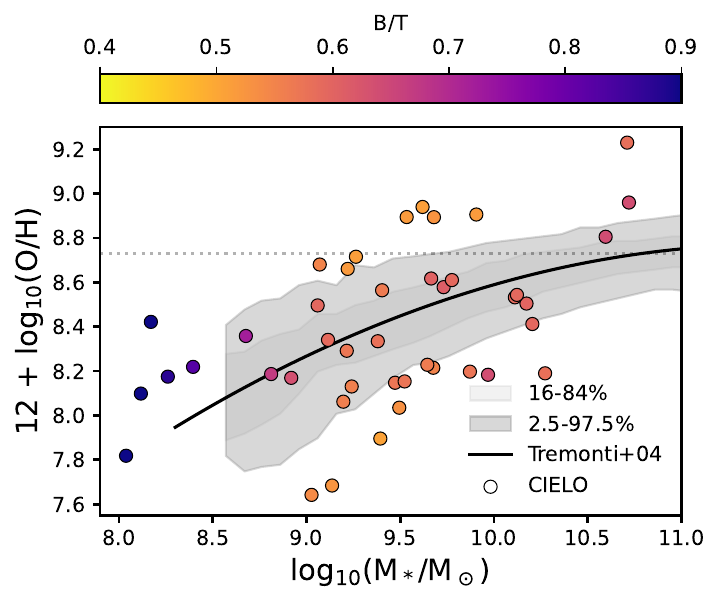}
\includegraphics[width=0.45\textwidth]{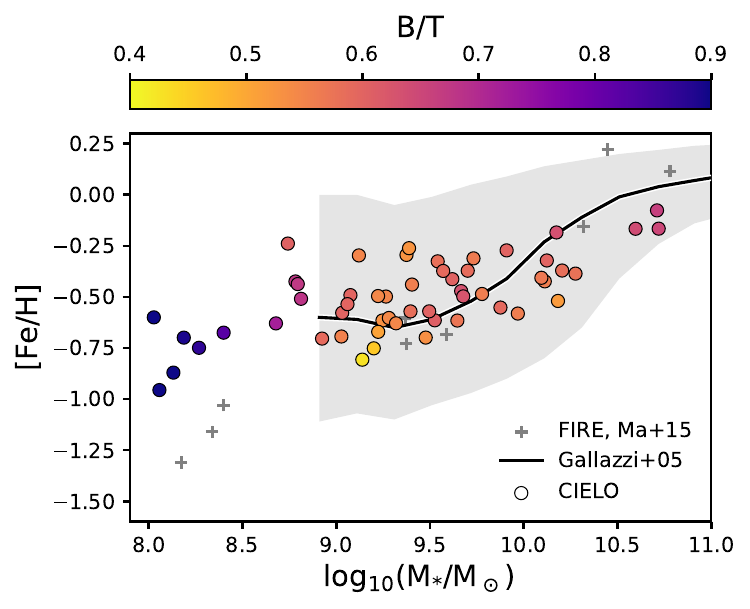}
\caption{Mass-metallicity relation for the star-forming gas in the ~\cielo galaxies, coloured by the B/T. Observations by \citet{tremonti2004} have been included for comparison rescaled to match the reported solar values by \citet[][black line]{asplund2009}. The shaded regions show the different percentiles as indicated by the labels. 
Lower panel: Mass-metallicity relation for the stellar populations of the \cielo~galaxies, coloured by the B/T. Results from FIRE simulations \citet{ma2015} and observations from \citet{gallazzi2005} are included for comparison (the shaded regions represent the 25-75$^{\rm th}$ percentiles).
}
\label{fig:mzr_SFR}
\end{figure}

 In the upper panel of Fig.~\ref{fig:mzr_SFR} the MZR for the star-forming gas in our simulated galaxies is displayed by using ${\rm 12 + log(O/H)}$. As can be seen the \cielo~ galaxies reproduce this relation fairly well at $z=0$ \citep{tremonti2004}. However, a significant dispersion appears to be present. 
To test and quantify the secondary dependence of metallicity on SFR, \reff~and B/T, we resorted to correlation matrix and random forest regressions.
To do this, we estimated the residuals with respect to the median MZR and searched for correlations of the residuals with B/T, SFR and \reff~ by using the Pearson correlation factor.
 We find  no clear  correlation between the residuals and \reff~, B/T or sSFR as can be seen in Fig.~\ref{fig:matrix-gas}.
   However, a random forest analysis provides signals of similar level of importance for the three parameters with a larger one being \reff~ as shown in the upper panel of Fig.~\ref{fig:randomSF}. The MSE and $R^2$ obtained are 0.15 and 0.35, consistent with a larger dispersion in the simulated data, although it globally follows the observed MZR\footnote{MSE is the mean square error and low value is preferable.  The coefficient of determination, $R^2$, measures the proportion of variance in the dependent variable that is predictable from the independent variables. A value close to unity is preferable.}.

The stellar populations in the \cielo~galaxies also show a MZR in global agreement with the observational results  \citep{gallazzi2005} and other simulations as can be seen  from the lower panel of  Fig.~\ref{fig:mzr_SFR}. In this case, we use [Fe/H] as an indicator of metallicity.   We  find a trend for  more disc-dominated systems to be be less enriched at a given \mstar. 
A similar analysis using the correlation matrix and random forest regressions was performed for the stellar MZR and the same parameters as in the case of the star-forming gas MZR. For the stellar populations, the residuals show an anticorrelation with SFR (${\rm -0.27, p < 0.001}$) and \reff~ (${\rm -0.51, p < 0.001 }$) and correlation for B/T (${\rm -0.19, p=0.07}$; see Fig.~\ref{fig:matrix-gas}).   The random forest analysis provides a larger level of importance to SFR followed by the \reff~ and the B/T as can be seen from the lower panel of Fig.~\ref{fig:randomSF}.  The stellar MZR is better defined and shows less dispersion which allows us to recover trends, which indicates that higher enriched galaxies have lower star formation activity, are more compact, and tend to be more dispersion-dominated. 

We find that the \cielo~ galaxies follow the expected trend, where more disc-dominated systems with larger sizes and higher star formation rates tend to exhibit lower levels of enrichment at a given stellar mass \citep{ellison2008, neumann2021}.

\section{Summary}

We present a new suite of chemo-dynamical zoom-in simulations from the \cielo~ project. The evolution of the simulated galaxies and their nearby environment is followed at the same level of resolution. The ICs   map groups, filaments, voids and walls. The analysed sample is restricted to a stellar mass range of $[10^8, 10^{11}] \Msun$. The current sample encompasses 54 central galaxies.

 We characterised the global environment of the \cielo~galaxies by applying DisPerSe and determined the position of the target galaxies in each simulated volume within the cosmic web. This was done only for the Pehuen haloes since the LG volumes were selected by using specific criteria of relative position and velocities between the two main galaxies. 

The morphology of the simulated galaxies was determined through a dynamical decomposition to identify their bulge, disc, and stellar halo components \citep{tissera2012}. This decomposition enables the estimation of morphology, which is found to correlate with galaxy size and to exhibit a secondary dependence on both the mass-size relation and the MZR.

Our main results are the followings:
\begin{itemize}
    \item The \cielo~galaxies are found to host stellar-mass halo fractions consistent with observations and other simulations. However, the sample size is still too small to perform a comprehensive analysis of potential environmental dependences.

    \item The \cielo~galaxies exhibit a star-formation activity slightly lower than expected at $z=0$ \citep{salim2018} . However, they have the expected stellar mass content per halo. This suggests that an overall modulation of star formation activity may be necessary, ensuring it does not disrupt the well-established fundamental relations. Additionally, we must consider that \cielo~galaxies are located in specific environments, where star formation activity might be influenced or even reduced, for example in filaments \citep[e.g.][]{okane2024}.

\item The mass-size relation shows a clear dependence on galaxy morphology, with more extended galaxies at a given \mstar~tending to be more disc-dominated, consistent with observational findings \citep[e.g.][]{vanderwel2014}.

\item The \cielo~galaxies follow the mass-size relation and the TFR, suggesting that the regulation of star formation activity and the implemented subgrid feedback have produced a sample of galaxies that, overall, closely resemble observations. Nevertheless, we find that our galaxies are slightly less star-forming, albeit still within the observed range. We also need to consider possible environmental effects.

\item The star-forming gas and stellar MZR are found to be in very good agreement with observations. The stellar MZR exhibits less dispersion and shows a clear secondary dependence on \reff, SFR, and B/T, with SFR appearing to have the strongest influence. These trends suggest that, at a given stellar mass, more compact galaxies have higher metallicities and are less star-forming, consistent with observational findings \citep[e.g.][]{ellison2018}.

\end{itemize}

In future works, we will investigate the properties and formation pathways of the bulge, disc, and stellar haloes, with a focus on their chemo-dynamical characteristics and their connections to assembly histories and the cosmic web. Additionally, we plan to continue running simulations for the Pehuen haloes to achieve more robust statistical results. This will enable a deeper understanding of galaxies forming in diverse environments such as voids, filaments, and walls, shedding light on the interplay between galaxy assembly and their surrounding environment \citep[e.g.][]{Dominguez2023b}.

\begin{acknowledgements}
    PBT acknowledges partial funding by Fondecyt-ANID 1240465/2024, N\'ucleo Milenio ERIS NCN2021\_017, and ANID Basal Project FB210003. 
    We acknowledge support from the European Research Executive Agency HORIZON-MSCA-2021-SE-01 Research and Innovation programme under the Marie Skłodowska-Curie grant agreement number 101086388 (LACEGAL).
              This project used the Ladgerda Cluster (Fondecyt 1200703/2020 hosted at the Institute for Astrophysics, Chile), the NLHPC (Centro de Modelamiento Matem\'atico, Chile) and the Barcelona Supercomputer Center (Spain).
\end{acknowledgements}

\bibliographystyle{aa} 
\bibliography{bibliography.bib}

\newpage

\begin{appendix}

\section{Numerical convergence}
\label{appendix}
\begin{figure}
\centering
\includegraphics[width=0.45\textwidth]{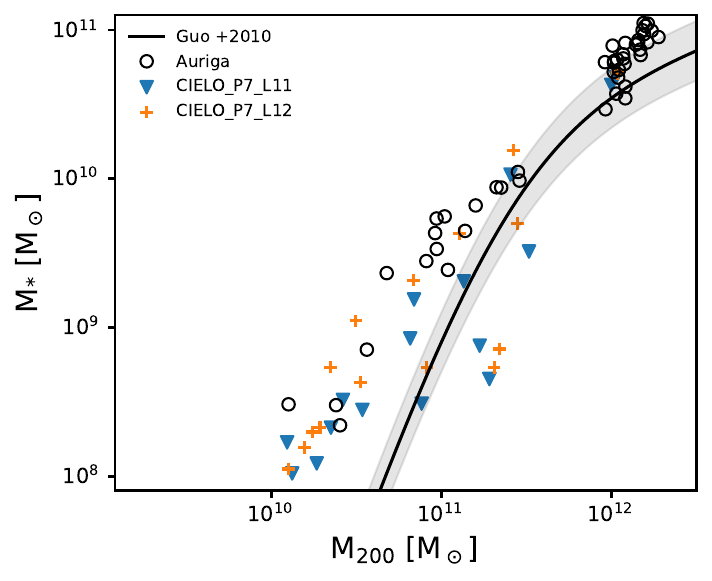}
\includegraphics[width=0.45\textwidth]{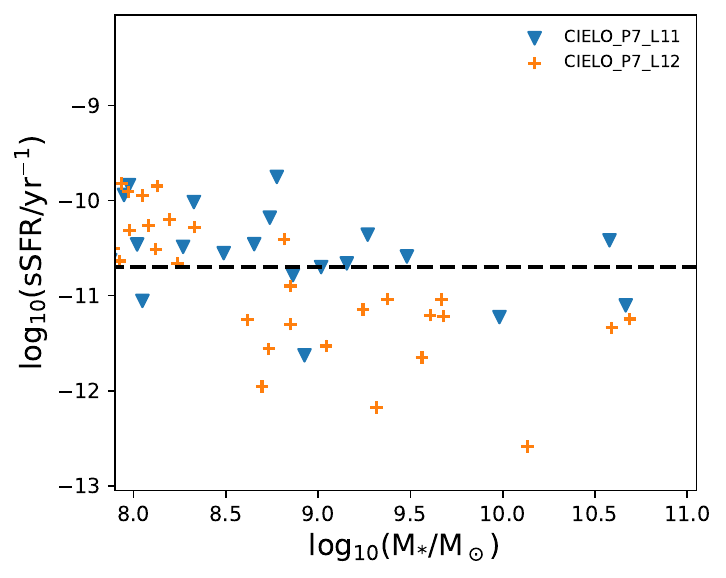}
\caption{Upper panel: Stellar mass versus virial halo mass for central galaxies in CIELO-P7 performed with two level of resolution: L11 and L12 (see inset labels). Lower panel: Specific SFR as a function of the stellar mass for the same galaxies. 
}
\label{fig:comparison}
\end{figure}

We run CIELO-P7 at L12 resolution level and hence, we can assess the impact on resolution on the simulated trend. In the upper panel of Fig.~\ref{fig:comparison} we show the \mstar~ as a function of \mvir for galaxies in both runs. They both follow the expected trend, but L12 run seems to have formed slightly more stars. From the lower panel of  Fig.~\ref{fig:comparison}, it can be seen that the higher resolution run have slightly less star-forming systems particularly at the high-mass end. We used the same parameters for both runs so this might be the consequence of the L12 achieving higher densities and forming slightly more stars at high-z.

\section{Estimation of correlations}
\begin{figure}
\resizebox{8cm}{!}{\includegraphics{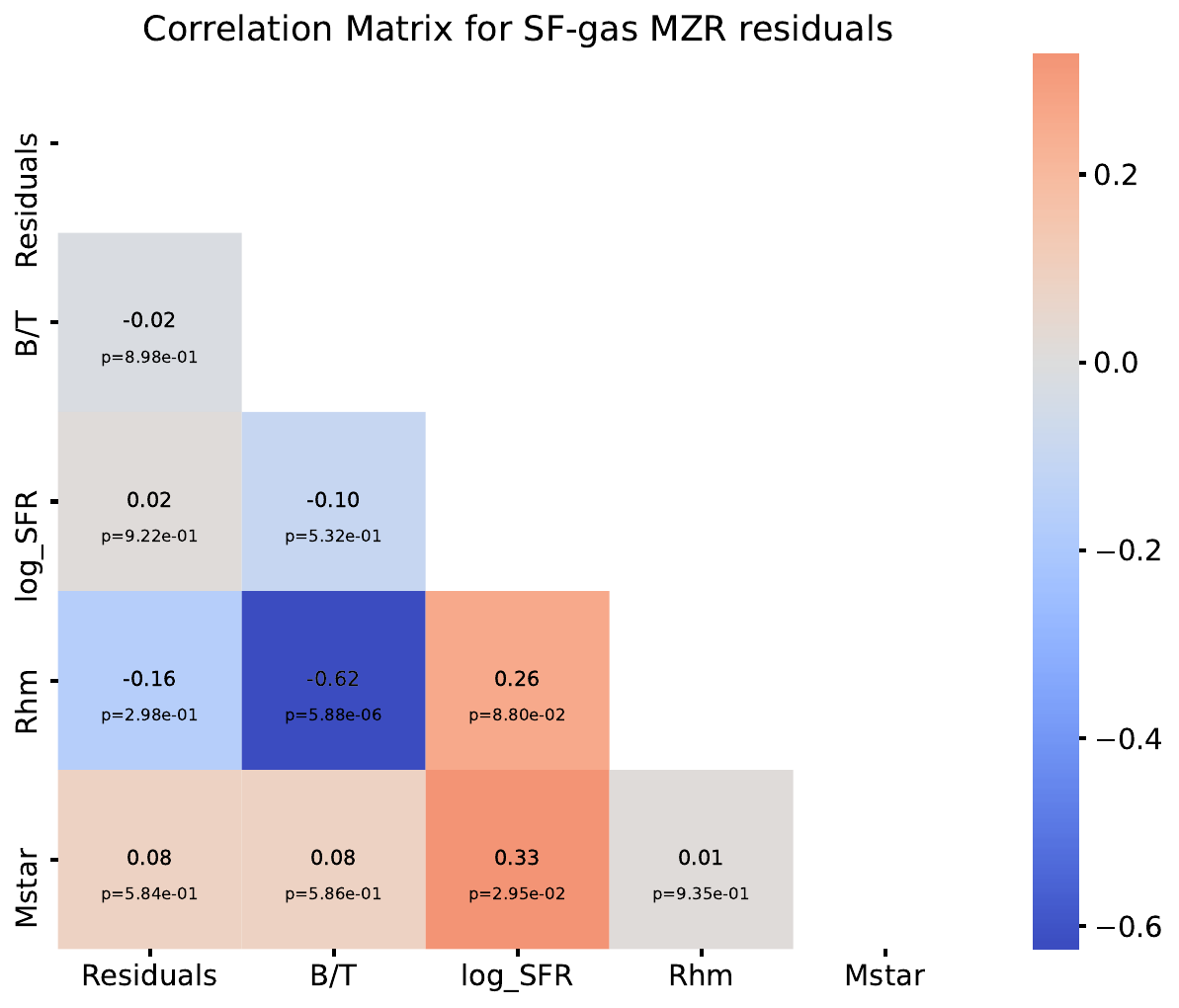}}
\resizebox{8cm}{!}{\includegraphics{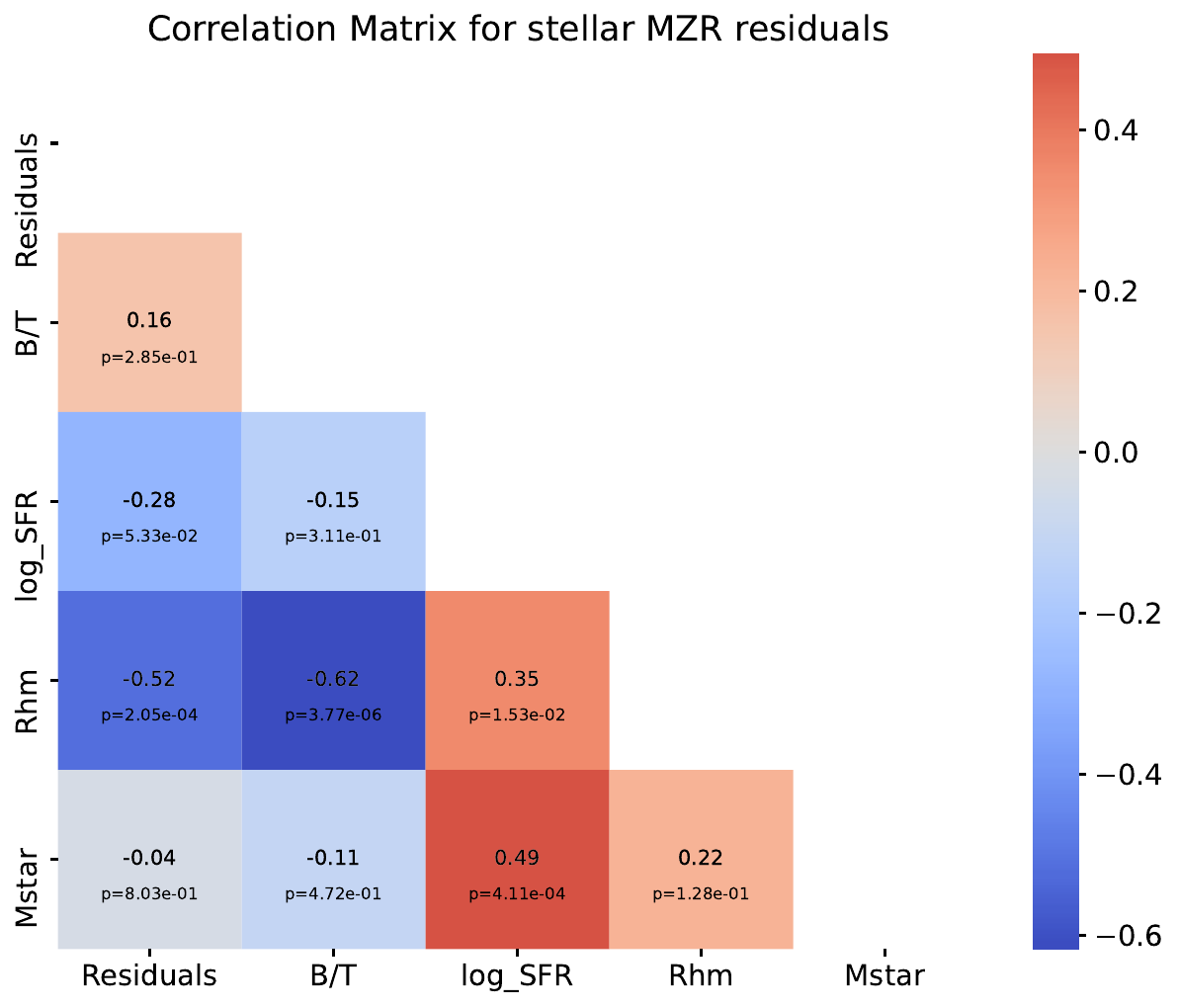}}
\caption{Correlation coefficient matrix illustrating the relationships between the residuals of the  star-forming (upper panel) and stellar (lower panel) MZR, and B/T, SFR, \mstar  and \reff. The matrix shows also the cross-correlated coefficients between these parameters.  
}
\label{fig:matrix-gas}
\end{figure}

\begin{figure}
\resizebox{8cm}{!}{\includegraphics{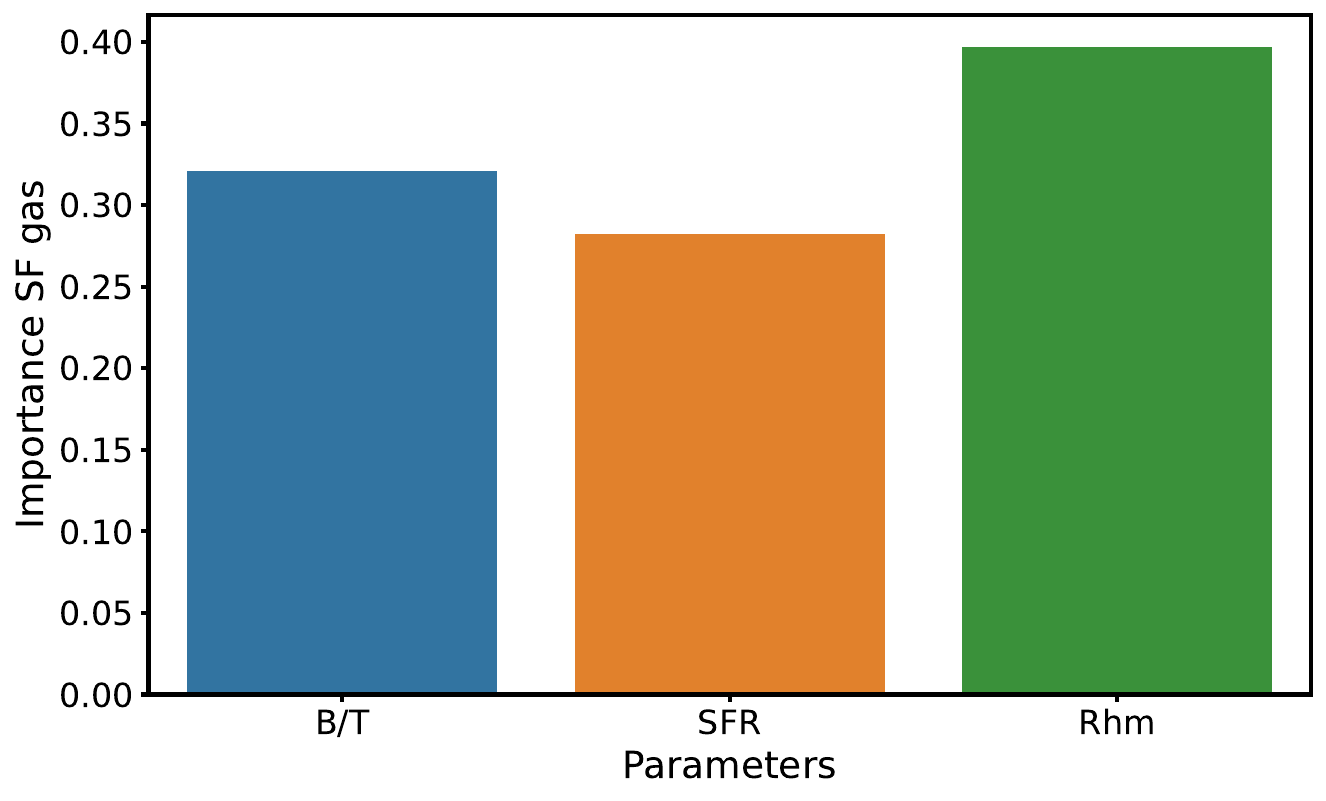}}
\resizebox{8cm}{!}{\includegraphics{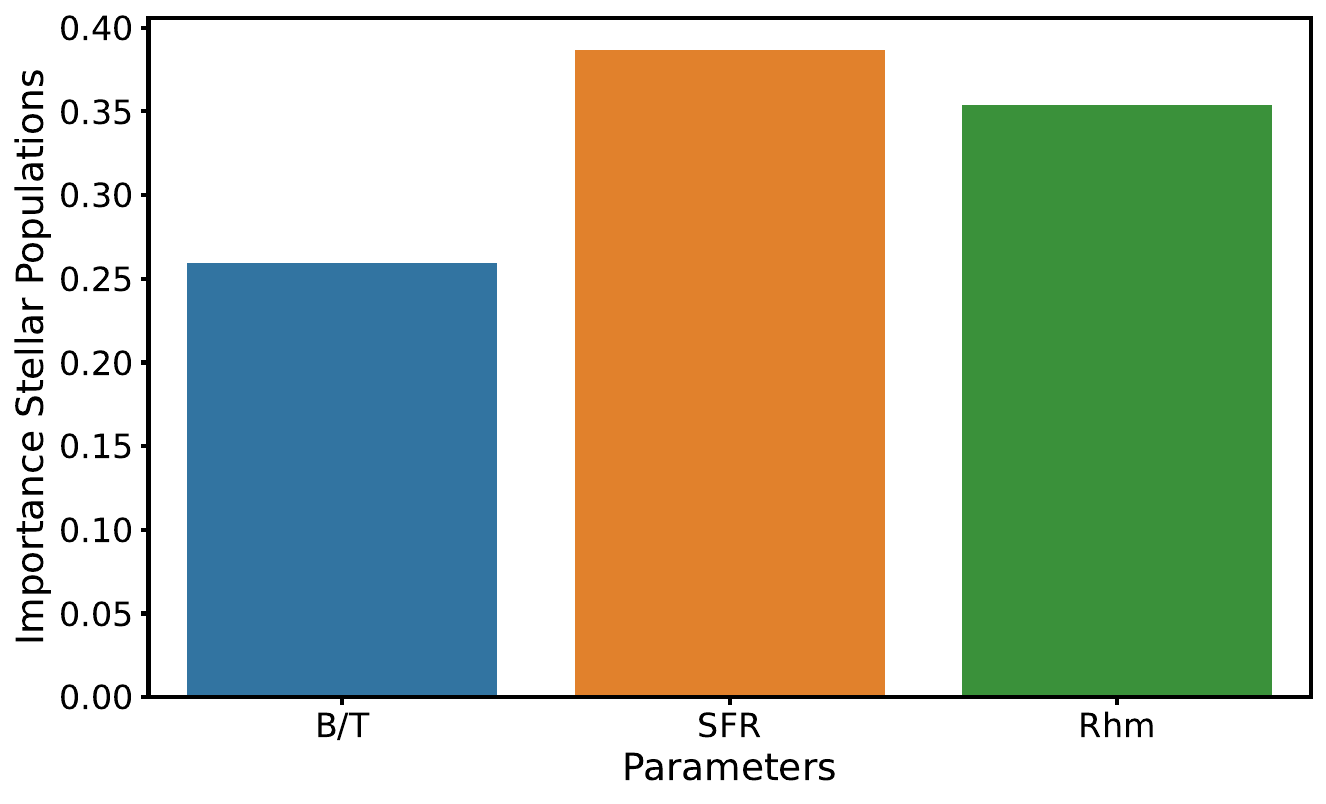}}
\caption{Random forest analysis between the residuals of the star-forming (upper panel) and stellar (lower panel) MZR and B/T, SFR and \reff.
}
\label{fig:randomSF}
\end{figure}

To assess secondary dependences and correlations between the properties of \cielo~galaxies and the residual of the MZR, we resort to the Pearson Correlation factors as shown by the correlation matrix for the residuals of the stellar MZR and star-forming gas MZR shown in Fig.\ref{fig:matrix-gas}.
Additionally, we applied  a random forest analysis to assess which of the three parameters exhibit a higher impact on both relations as displayed in Fig.~\ref{fig:randomSF}.

\end{appendix}

\end{document}